\documentclass[12pt]{iopart}

\usepackage{iopams}  

\usepackage{epsfig}
\usepackage{graphicx}
\usepackage[figuresright]{rotating}

\newcommand{\bea}{\begin{eqnarray}}
\newcommand{\eea}{\end{eqnarray}}
\newcommand{\bean}{\begin{eqnarray*}}
\newcommand{\eean}{\end{eqnarray*}}
\newcommand{\nn}{\nonumber \\}

\def\IC{\mathbb{C}}
\def\IP{\mathbb{P}}

\def\W #1{\widetilde{#1}}

\def\ket#1{\left| #1\right\rangle}

\def\gb #1{ \left\langle #1 \right]}
\def\tgb #1{ \left[ #1 \right\rangle}
\def\cb #1{ \left[ #1 \right]}

\def\bbar#1{ \overline #1}

\def\det{\mathop{\rm det}}

\def\la{\lambda}
\def\tl{\tilde\lambda}
\def\lt{\tilde\lambda}
\def\eps{\epsilon}

\def\vev#1{\left\langle #1 \right\rangle}

\def\half{\frac{1}{2}}
\def\ord{{\cal O}}
\def\N{{\cal N}}

\def\Label#1{\label{#1}}

\begin{document}

\title[Loop amplitudes in gauge theories: modern analytic approaches]{Loop amplitudes in gauge theories: modern analytic approaches}

\author{Ruth Britto}

\address{IPhT, CEA-Saclay, 91191 Gif-sur-Yvette cedex, France}

\begin{abstract}
This article  reviews  on-shell methods for analytic computation of loop amplitudes, emphasizing techniques based on unitarity cuts.  
Unitarity techniques are formulated generally but have been especially useful for calculating one-loop amplitudes in massless theories such as  Yang-Mills theory, QCD, and QED.

\end{abstract}

\maketitle

\section{Introduction}

Scattering amplitudes can be constructed in terms of their singularities. For tree amplitudes, these are complex poles.  In loop amplitudes, there are branch cuts, as well as other singularities associated with ``generalized'' cuts.  All of these singularities probe factorization limits of the amplitude:  they  select kinematics where some propagators are put {\em on shell}.  Thus, the calculation can be packaged in terms of lower-order {\em amplitudes} instead of the complete sum of Feynman diagrams.

The ``unitarity method'' started as a framework for one-loop calculations.  Instead of the explicit set of loop Feynman diagrams, the basic reference point is the linear expansion of the amplitude function in a basis of ``master integrals,'' multiplied by coefficients that are rational functions of the kinematic variables.  The point is that the most difficult part of the calculation, namely integration over the loop momentum, can be done once and for all, with explicit evaluations of the master integrals.  The master integrals contain all the logarithmic functions.  It then remains  to find their coefficients.

If an amplitude is uniquely determined by its branch cuts, it is said to be {\em cut-constructible}.   
All one-loop amplitudes are cut-constructible in dimensional regularization, provided that the full dimensional dependence is kept in evaluating the branch cut.
Each master integral has a distinct branch cut, uniquely identified by its logarithmic arguments.  Therefore, the decomposition in master integrals can be used to solve for their coefficients separately using analytic properties.  It is not necessary to reconstruct the amplitude from the cut in a traditional way from a dispersion integral.  Rather, we overlay information from various cuts separately.

Beyond one-loop order, not many analytic results have been obtained by unitarity methods in the absence of maximal supersymmetry.  
Perhaps the main obstacle is the lack of a manageable, systematic, explicitly known set of master integrals.  
Also, $D$-dimensional unitarity cuts of higher-loop amplitudes  
involve lower-order amplitudes which still contain loops and yet have $D$-dimensional momenta on some external legs (the cut lines).

Analytic calculations are simplest in massless theories, where formulas can be written compactly in the spinor-helicity formalism.  Spinor variables are helpful inside the loop as well, when propagators associated to massless field are placed on shell in a unitarity cut.  For this reason as well, we work mostly  in four-dimensional Minkowski space and its analytic continuations.

This article is organized as follows.  In Section 2, we introduce the master integrals and review integral reduction.  In Section 3, we describe the unitarity method and cut integrals.  We evaluate the cuts of master integrals, explain the evaluation of general cut amplitudes, and list formulas for the coefficients of of master integrals, given a general one-loop integrand. In Section 4, we discuss generalized unitarity cuts for one-loop amplitudes, from quadruple and triple cuts to single cuts along with adaptations of Feynman's Tree Theorem.  In Section 5, we address $D$-dimensional unitarity methods, which can be used either for an exact expression of the regulated amplitude or to avoid the separate calculation of ``rational terms'' in addition to the master integral coefficients.  In Section 6, we mention issues that arise in the presence of massive particles.  
In Section 7, we review applications of MHV-diagram constructions.  In Section 8, we discuss on-shell recursion relations for coefficients or rational terms in one-loop amplitudes.  In section 9, we summarize recently published analytic results for one-loop amplitudes.  Finally, in section 10, we look at prospects for continuing progress along these lines to higher-loop amplitudes.

\section{One-Loop Master Integrals}

One-loop Feynman integrals can be expressed as a linear combination of master integrals with rational coefficients.  The master integrals are typically taken to be the set of scalar integrals, meaning that there is no tensor structure left in the numerator.  The scalar $n$-point integral is defined as
\bean
\hspace{-1in}
I_n = (-1)^{n+1} i (4\pi)^{\frac{D}{2}} 
\int \frac{d^{D}\ell}{(2\pi)^{D}}
\frac{1}{(\ell^2-m_1^2)((\ell-K_1)^2-m_2^2)((\ell-K_1-K_2)^2-m_3^2)\cdots
((\ell+K_n)^2-m_n^2)}
\Label{master-def}
\eean
Here, the $K_i$ are sums of external momenta.  They are strictly four-dimensional.

\begin{figure}
\includegraphics[width=15cm]{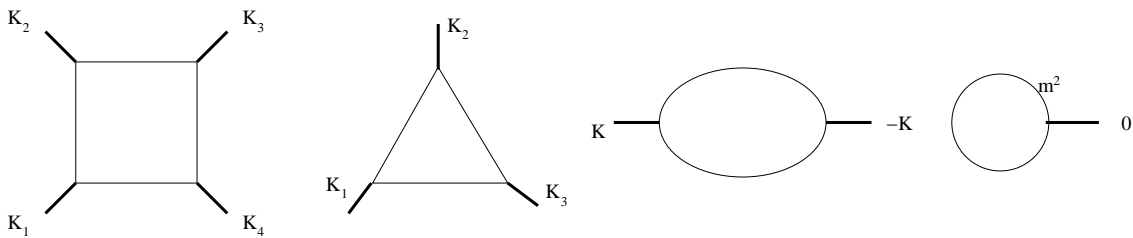}
\caption{One-loop master integrals: box, triangle, bubble and tadpole.  The Lorentz vectors $K_i$ are sums of external momenta, all directed outward.}
\end{figure}

Working in theories with mostly massless particles, it is natural to use dimensional regularization, i.e.  $D=4-2\eps$.  We are often satisfied with a result good to $\ord(\eps)$.  In analytic computations, we keep the external momenta fixed in four dimensions in order to use the spinor-helicity formalism.

Integral reduction \cite{Brown:1952eu, Melrose:1965kb, Passarino:1978jh, Hooft1978xw, vanNeerven:1983vr, Bern:1992em, Bern:1993kr} is a clearly defined procedure for expressing any one-loop Feynman integral as a linear combination of scalar boxes, scalar triangles, scalar bubbles, and scalar tadpoles, with rational coefficients:  
\bea
A^{\rm 1-loop} = \sum_n \sum_{{\bf K}=\{K_1,\ldots,K_n\}} c_n({\bf K})  I_n({\bf K})
\Label{basis-exp}
\eea
In four dimensions, $n$ ranges from 1 to 4.  In dimensional regularization, the tadpole contributions with $n=1$ arise only with internal masses.  If we keep higher order contributions in $\eps$, we find that the pentagons ($n=5$) are independent as well.

Let us briefly outline the traditional reduction procedure.  We assume that the integral has been constructed from Feynman diagrams, so that the denominators are propagators of the form $D_i = (\ell-P_i)^2-M_i^2$, along with the propagator defining the loop momentum, $D_0 = \ell^2-M_0^2$, chosen anew for each term at each stage.
There are three steps.  

First, we eliminate tensor structure (i.e. momentum-dependent numerators) in terms with at least five propagators.  Any appearance of $\ell^2$ in the numerator is replaced by $M_0^2+D_0$, and the $D_0$ term cancels against the denominator.  The remaining momentum dependence in the numerator is polynomial in contractions of the form $\ell\cdot P$.
Among the five propagators, there are four independent momentum vectors $P_i$ in which to expand any $P$.  Then we make the replacement $2\ell\cdot P_i = M_o^2+P_i^2-M_i^2+D_0-D_i$ and cancel $D_0$ and $D_i$ against the denominator.  Step by step, the degree of the polynomial is lowered until we have a scalar numerator or at most four propagators in the denominator.

Second, we eliminate remaining tensor structure in the terms with at most four propagators.  This is done by using the momenta appearing in the denominators to build a basis of Lorentz-covariant tensors in which to expand the integral.  Contracting the tensors with external momenta gives the constraints needed to solve the linear system.  It can be particularly efficient to use contractions with complex momenta constructed from spinors associated to different external legs \cite{Pittau:1996ez,Pittau:1997mv,Weinzierl:1998we,delAguila:2004nf,Pittau:2004bc}.

Third, we express $n$-point scalar integrals with $n>4$ in terms of lower-point scalar integrals.  If $n \geq 6$, then there is a nontrivial solution $\{\alpha_i\}$ to the five equations $\sum_{i=1}^n \alpha_i=0$ and $\sum_{i=1}^n \alpha_i P^\mu_i=0$.  With this solution, $\sum_i \alpha_i D_i = \sum_i \alpha_i (P_i^2-M_i^2)$.  Divide the integrand by the (momentum-independent) right-hand side of this equation and multiply it by the left-hand side.  The factors $D_i$ will cancel against the denominator and reduce $n$ by one. The final remaining concern is the scalar pentagon.  If we are keeping full $\eps$ dependence in dimensional regularization, it is an independent master integral.  
If we truncate the integrals at $\ord(\eps)$, then the scalar pentagon can be reduced further to four scalar boxes.  The scalar pentagon integral is finite, so we can now treat its loop momentum as four-dimensional.  The final reduction involves expanding it in terms of the axial vectors constructed from triples of four independent external momenta.

Explicit formulas for the divergent one-loop master integrals are given in \cite{Ellis:2007qk}, for all arrangements of massive and massless lines.   In the next few sections, we will study unitarity cuts specifically at the case in which all internal lines are massless.  Then the tadpoles are absent.  We now list the necessary master integrals in this case evaluated through $\ord(\eps)$.  The expressions here are taken from \cite{Bern:1994cg,Denner:1991qq}.  Other useful expressions for scalar box integrals, convenient for analytic continuation to different kinematic regions, appear in \cite{Duplancic:2000sk}.

The dimensional regularization parameter is $\epsilon=(4-D)/2$.  The constant $r_\Gamma$ is defined by
\bea
r_\Gamma = {\Gamma(1+\epsilon)\Gamma^2(1-\epsilon) \over \Gamma(1-2\epsilon)}
\label{rgamma}
\eea

\noindent{\bf Scalar bubble integral, no internal masses:} 
\bea
I_{2}=r_{\Gamma}
\left({1\over \epsilon} -\ln(-K^2)+2\right) +\ord(\epsilon)
\Label{masterbubble}
\eea

\noindent {\bf Scalar triangle integrals, no internal masses:} \\
If $K_2^2=K_3^2=0$ and $K_1^2 \neq 0$, then the scalar triangle is called ``one-mass'', and it is
\bea
I^{1m}_{3}& =& { r_{\Gamma} \over \epsilon^2} (-K_1^2)^{-1-
\epsilon}.
\eea
If $K_3^2=0$ and $K_1^2,K_2^2 \neq 0$, then the scalar triangle is called ``two-mass'', and it is
\bea
I^{2m}_{3} & =&  { r_{\Gamma} \over \epsilon^2} {
(-K_1^2)^{-\epsilon} -(-K_2^2)^{-\epsilon}
\over (-K_1^2)-(-K_2^2)}
\eea
The ``three-mass'' scalar triangle is finite and given by
\bea
I^{3m}_{3} & =&  {i \over \sqrt{\Delta_3}} \sum_{j=1}^3
\left[ {\rm Li}_2\left(-{1+i\delta_j \over 1-i \delta_j} \right)
-{\rm Li}_2\left(-{1-i\delta_j \over 1+i \delta_j} \right)
\right]+\ord(\epsilon),
\eea
where we have defined the following:
\bea
& & \Delta_3 =  -(K_1^2)^2-(K_2^2)^2-(K_3^2)^2+ 2K_1^2 K_2^2
+2 K_2^2 K_3^2 +2 K_3^2 K_1^2 
\Label{tri-kallen}
\\
& & \delta_j =  {2K_j^2 -(K_1^2+K_2^2+K_3^2)
\over \sqrt{\Delta_3}} 
\eea

\noindent {\bf Scalar box integrals, no internal masses:}\\ 
Let $s=(K_1+K_2)^2$ and $t=(K_1+K_4)^2$.  The dilogarithm function is defined by ${\rm Li}_2(x) = -\int_0^x \ln
(1-z)dz/z$.

If all four momenta are massless, i.e. $K_1^2=K_2^2=K_3^2=K_4^2=0$ (a special case for four-point amplitudes), then the box integral is given by
\bea
 I^{0m}_{4} & =& {r_{\Gamma} \over st}\left({2\over \epsilon^2}
\left[ (-s)^{-\epsilon} +(-t)^{-\epsilon} \right]-
\ln^2\left({s\over t} \right) -\pi^2
\right)+\ord(\epsilon).
\eea
If only one of the four momenta, say $K_1$, is massive, and the other are massless, i.e. $K_2^2=K_3^2=K_4^2=0$, then the box is called ``one-mass'', and it is given by
\bea
 I^{1m}_{4} & =&
{2r_{\Gamma} \over st} ~{1\over \epsilon^2}\left[
(-s)^{-\epsilon} +(-t)^{-\epsilon}
-(-K_1^2)^{-\epsilon} \right]  \\
& & \nonumber
 -{2r_{\Gamma} \over st}
\left[  {\rm Li}_2 \left( 1-
{K_1^2\over s} \right) + {\rm Li}_2 \left( 1-
{K_1^2 \over t} \right) + {1\over 2}\ln^2\left(
{s\over t} \right) + {\pi^2\over 6}
\right]+\ord(\epsilon).
\eea
There are two distinct arrangements of two massive and two massless legs on the corners of a box.   
In the ``two-mass-easy'' box, the massless legs are diagonally opposite.  If $K_2^2=K_4^2=0$ while the other two legs are massive, the integral is 
\bea
 I^{2m~e}_{4} & =&
{2r_{\Gamma} \over st-K_1^2K_3^2}
 ~ {1\over
\epsilon^2}\left[ (-s)^{-\epsilon}
+(-t)^{-\epsilon} -(-K_1^2)^{-\epsilon} -
(-K_3^2)^{-\epsilon} \right] 
\\ & & \nonumber
 -{2r_{\Gamma} \over st-K_1^2K_3^2}
\left[ {\rm Li}_2 \left(
1- {K_1^2\over s} \right) + {\rm Li}_2 \left(
1- {K_1^2\over t} \right) +{\rm Li}_2 \left( 1-
{K_3^2\over s} \right)\right. 
\\ & & \nonumber
+ \left.{\rm Li}_2
\left( 1- {K_3^2\over t} \right) - {\rm
Li}_2 \left( 1- {K_1^2 K_3^2\over
st } \right) + {1\over 2}\ln^2\left(
{s\over t} \right) \right] + \ord(\epsilon).
\eea
In the ``two-mass-hard'' box, the massless legs are adjacent.  If $K_3^2=K_4^2=0$ while the other two legs are massive, the integral is 
\bea
 I^{2m~h}_{4} &=&
{2r_{\Gamma} \over st}
~ {1\over
\epsilon^2}\left[ \half (-s)^{-\epsilon}
+(-t)^{-\epsilon} -\half (-K_1^2)^{-\epsilon} -
\half (-K_2^2)^{-\epsilon} \right] 
\\ & & \nonumber
-{2r_{\Gamma} \over st}
\left[ -\half \ln\left({s \over K_1^2}\right)
\ln\left({s \over K_2^2 }\right)
+ {1\over 2}\ln^2 \left(
{s\over t} \right)\right. 
\\ & & \nonumber
\left. +  {\rm Li}_2
\left( 1- {K_1^2\over t} \right) + {\rm Li}_2
\left( 1- {K_2^2\over t} \right) \right] +\ord(\eps). 
\eea
If exactly one leg is massless, say $K_4^2=0$, then we have the ``three-mass'' box, given by
\bea
 I^{3m}_{4} &=&
{2r_{\Gamma} \over st-K_1^2 K_3^2} 
~ {1\over \epsilon^2}
\left[\half (-s)^{-\epsilon}
+\half (-t)^{-\epsilon} -\half (-K_1^2)^{-\epsilon}
- \half (-K_3^2)^{-\epsilon}
\right]
\\ & & \nonumber
-{2r_{\Gamma} \over  st-K_1^2 K_3^2}
\left[
-\half \ln\left({s \over K_1^2 }\right)
\ln\left({s \over K_2^2 }\right)
-\half \ln\left({t \over K_2^2}\right)
\ln\left({t \over K_3^2 }\right)
\right. 
\\ & & \nonumber
\left. {1\over 2}\ln^2 \left(
{s \over t } \right)+  {\rm Li}_2
\left( 1- {K_1^2 \over t} \right) + {\rm Li}_2
\left( 1- {K_3^2 \over s} \right) - {\rm
Li}_2 \left( 1- { K_1^2 K_3^2 \over
st } \right) \right] + \ord(\eps). 
\eea
Finally, the ``four-mass'' box, which is finite, is given by
\bean
I^{4m}_4 &=& {1\over a (x_1-x_2)}\sum_{j=1}^2 (-1)^j
\left( -\half \ln^2(-x_j) \right. \\ 
&&   -{\rm Li}_2\left( 1+
{-K_3^2-i\varepsilon \over -s-i\varepsilon }x_j \right)
-\eta\left( -x_k, {-K_3^2-i\varepsilon \over -s-i\varepsilon }
\right) \ln \left( 1+ {-K_3^2-i\varepsilon \over -s-i\varepsilon
}x_j \right) \\
& & -{\rm Li}_2\left( 1+ {-t-i\varepsilon \over
-K_1^2-i\varepsilon }x_j \right) -\eta\left( -x_k,
{-t-i\varepsilon \over -K_1^2-i\varepsilon } \right) \ln \left( 1+
{-t-i\varepsilon \over -K_1^2-i\varepsilon }x_j \right) \\
& &
\left. +\ln(-x_j)(\ln(-K_1^2-i\varepsilon ) + \ln(-s-i\varepsilon
) - \ln(-K_4^2-i\varepsilon ) - \ln(-K_2^2-i\varepsilon )  ) \right). 
\eean
Here we have defined 
\bean
\eta(x,y)=2\pi i [\vartheta(-{\rm Im~} x)\vartheta(-{\rm Im~}
y)\vartheta({\rm Im~}(xy)) -\vartheta({\rm Im~} x)\vartheta({\rm Im~}
y)\vartheta(-{\rm Im~}(xy))], 
\eean
and $x_1$ and $x_2$ are the roots of a quadratic polynomial:
\bea
a x^2+b x+ c + i\varepsilon d = a (x-x_1)(x-x_2),
\Label{quadratic}
\eea
with
\bean
 a = t K_3^2, \quad
 b= st + K_1^2 K_3^2 - K_2^2 K_4^2, \quad
 c = s K_1^2, \quad
 d = -K_2^2. 
\eean

\subsection{Simple theories}

Some field theories are known to obey additional constraints on their expansion in master integrals. 
Maximally supersymmetric theories, $\N=4$ SYM and $\N=8$ supergravity, have a ``no-triangle'' property \cite{Bern:1994zx,Bern:2005hh,BjerrumBohr:2006yw,BjerrumBohr:2008vc,BjerrumBohr:2008ji,ArkaniHamed:2008gz}.  One-loop amplitudes can be written in terms of box integrals only, without triangles or bubbles.  This property follows from supersymmetric cancellations combined with power-counting, although for supergravity, new reduction formulas needed to be found.  Master integrals for higher-loop amplitudes should not contain triangle or bubble subdiagrams, either.  

Massless supersymmetric theories should be four-dimensionally cut-constructible.  That is, the expansion in master integrals (\ref{basis-exp}) is valid when truncated at $\ord(\eps)$, without any additional rational terms that could arise from the expansion in $\eps$. This was shown specifically for color-ordered gauge theory amplitudes \cite{Bern:1994zx,Bern:1994cg} but is expected to be more generally valid, and is also useful in organizing calculations in nonsupersymmetric theories.   

Certain gauge theories with reduced supersymmetry still feature enough symmetry to guarantee the cancellation of triangle or bubble contributions.  
In $\N=6$ supergravity, bubble contributions are absent at one loop \cite{Dunbar:2010fy}.
In Yang-Mills, theories with triangle or bubble cancellations have been recently characterized in terms of the representation of matter content in \cite{Lal:2009gn}, whose conclusions we list here in Table \ref{simpleconds}.  An example of a theory with box integrals only is $\N=2$ with one hypermultiplet transforming in the symmetric tensor representation of the gauge group $SU(N)$  (where $N \geq 3$) and another hypermultiplet transforming in the antisymmetric tensor representation.  An example of a theory without bubbles is $\N=2$ with an $SU(N)$ gauge group and $2N$ hypermultiplets.
\begin{table}
\begin{center}
\label{simpleconds}
\begin{tabular}{|c|c|c|}\hline
\multicolumn{3}{|l|}{Condition ({\bf C}): ${\rm Tr}_{\rm R}(\Pi_{i=1}^n T^{a_i}) =  m\,{\rm Tr}_{\rm adj}(\Pi_{i=1}^n T^{a_i}),~ n \leq p$} \\ \hline
Non-susy theories have&only boxes & no bubbles \\ \hline
if $R_f$ satisfies {\bf C} with&p=6, m=4&p=4,m=4\\\hline
and $R_s$ satisfies {\bf C} with&p=6, m=6&p=4,m=6.\\\hline
Susy theories have&only boxes & no bubbles \\ \hline
if $R_{\chi}$ satisfies {\bf C} with&p=5, m=3&p=2,m=3.\\\hline
\end{tabular}
\caption{Conditions for cancellations of bubbles and triangles.  Matter representations are denoted by $f$ for fermions, $s$ for scalar and $\chi$ for chiral supermultiplets.  Taken from \cite{Lal:2009gn}.}
\end{center}
\end{table}

Multi-photon amplitudes in QED at one loop exhibit simplicity as well.  
Furry's theorem implies that the $n$-photon amplitude vanishes if $n$ is odd. 
Regarding the integral basis, it has been shown \cite{Badger:2008rn} that bubble integrals are absent for $n>4$, and both bubble and triangle integrals are absent for $n>6$.  Rational terms appear only for $n=4$.

\section{Unitarity methods\label{sec:umethod}}

The unitarity cut of a one-loop amplitude is its discontinuity across the branch cut in a kinematic region associated to a particular momentum channel.  The name comes from the unitarity of the S-matrix:  since $S^\dagger S=1$, and we expand $S=1+i T$ where $T$ is the interaction matrix, then $2 {\rm Im ~} T = T^\dagger T$.  Expanding this equation perturbatively in the coupling constant, we see that the imaginary part of the one-loop amplitude is related to a product of two tree-level amplitudes.  Effectively, two propagators within the loop are restricted to their mass shells.  This imaginary part should be viewed more generally as a discontinuity across a branch cut singularity of the amplitude---in a kinematic configuration where one kinematic invariant, say $K^2$, is positive, while all others are negative.  This condition isolates the momentum channel $K$ of interest; $K$ is the sum of some of the external momenta.  We will take cuts in various momentum channels to construct the amplitude.

For a one-loop amplitude, the  value of the unitarity cut is given by Cutkosky rules \cite{Cutkosky:1960sp}.\footnote{This technique and related approaches are discussed in \cite{sbook} in their original context, which excluded massless theories.  Modern interpretations were introduced in \cite{Bern:1994zx,Bern:1994cg}.}
The Cutkosky rules are expressed in the cut integral,
\bea
 \Delta A^{\rm 1-loop} \equiv \int d\mu~~ A^{\rm tree}_{\rm Left} ~\times~
 A^{\rm tree}_{\rm Right},
\Label{cutdef}
\eea
where the Lorentz-invariant phase space (LIPS) measure is defined by
\bea
d\mu = d^4{\ell_1}~ d^4{\ell_2}~ \delta^{(4)}(\ell_1+\ell_2 -
K)~ \delta^{(+)}(\ell_1^2)~\delta^{(+)}(\ell_2^2).
\Label{lips}
\eea
Here, the superscript $(+)$ on the delta functions for the cut propagators 
denotes the choice of a positive-energy solution.
\begin{figure}
\begin{center}
\includegraphics[width=7cm]{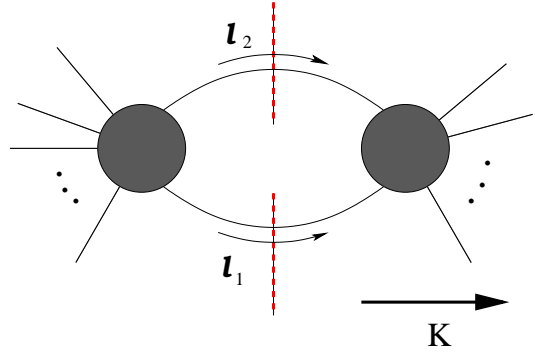}
\caption{Unitarity cut of a one-loop amplitude in the $K$ momentum channel.  The two propagators are constrained to their respective mass shells.  The disks represent the sum of all Feynman diagrams linking the fixed external lines and the two cut propagators.}
\end{center}
\end{figure}

How can these unitarity cuts be used to calculate the amplitude?  By applying the cut $\Delta$ in various momentum channels, we get information about the coefficients of master integrals.

Consider applying a unitarity cut to the expansion (\ref{basis-exp}) of an amplitude in master integrals.  Since the coefficients are rational functions, the branch cuts are located only in the master integrals.  Thus we find that 
\bea
\Delta A^{\rm 1-loop} = \sum_n \sum_{{\bf K}=\{K_1,\ldots,K_n\}} c_n({\bf K})  \Delta I_n({\bf K})
\Label{cut-exp}
\eea

Equation (\ref{cut-exp}) is the key to the unitarity method.  It has two important features.  First, 
we see from (\ref{cutdef}) that it is a relation involving {\em
  tree-level} quantities.  Second, many of the terms on the
right-hand side vanish, because only a subset of master integrals have
a cut involving the given momentum $K$.  Meanwhile, we enjoy the freedom of using all possible values of $K$ in turn.  In effect, we have traded the original single equation (\ref{basis-exp}) for a system of several shorter equations.

Our remaining task is to isolate the individual coefficients $c_i$.  With generalized unitarity, quadruple cuts isolate the box coefficients.  We have to work a little harder to get the triangle and bubble coefficients.  In the following section we will discuss triple cuts for triangle coefficients.  Here, we recall that four-dimensional cut-constructibility allows us to distinguish box, triangle, and bubble contributions from standard unitarity cuts, and we will see how these coefficients can be calculated in practice.

\subsection{Cuts of master integrals}

The utility of equation (\ref{cutdef}) depends on knowing the  master integrals and hence their branch cuts.  Having listed the master integrals for massless theories in the previous section, we can calculate the cuts explicitly by taking the imaginary parts of these functions in various kinematic regions.  Recall that the kinematic region associated to the unitarity cut in momentum channel $K$ has $K^2>0$ and all other invariants negative.  In the limit of large $K^2$, we see that the behavior of the master integrals features uniquely identifiable products of logarithms \cite{Bern:1994cg}.  It follows that no linear combination of master integrals with rational coefficients, in a given momentum channel, can be cut-free. 

The cuts of the bubble integrals are purely rational; this is easily seen from the ordinary logarithm in equation (\ref{masterbubble}).  The cuts of all other master integrals are logarithmic.  The various arguments of the logarithms identify the original master integrals.
 
We need to evaluate the left hand side of equation (\ref{cut-exp}), by carrying out the 2-dimensional integral of (\ref{cutdef}).  This is neatly accomplished by the Cauchy residue theorem, in a technique known as ``spinor integration'' \cite{Britto:2005ha,Britto:2006sj}.  We now illustrate the technique applied to the master integrals themselves, starting from their definition in (\ref{master-def}).

To implement the cut conditions, it is convenient to reparametrize the loop momentum in terms of spinor variables.  Now it is crucial that our cut is in 4 dimensions.  (The $D$-dimensional generalization will be presented in Section \ref{sec:ddim}.)

Since $\ell_1$ is null, we can parametrize it with 
\bea
(\ell_1)_{a\dot a}=t\la_a\tl_{\dot a},
\eea
where $\la_a,\tl_{\dot
  a}$ are {\em homogeneous} spinors (taking values in $\IC \IP^1$), and $t$ takes nonnegative values.  The original loop momentum is real-valued, so we will integrate over the contour where $\tl$ is the complex conjugate of $\la$ .
In the integral measure, we make the replacement\cite{Cachazo:2004kj}\footnote{The factor of $-4$ has been omitted in much of the literature, but such constant factors cancel out in unitarity methods and can be disregarded.}
\bea
\int d^4\ell_1~ \delta^{(+)}(\ell_1^2) (\bullet)
= - \int_0^\infty \frac{t}{4} dt \int_{\bbar\la=\tl} \vev{\la~d\la} \cb{\tl~d\tl}  (\bullet).
\eea 
Now we make this substitution explicitly in the second delta function of the LIPS measure defined in (\ref{lips}).  The momentum of the second cut propagator
is $(\ell_2)_{a\dot a}=K_{a\dot a}-t\la_a\tl_{\dot a}$, so the measure becomes 
\bea
 \int d\mu~~ (\bullet )  =
- \int_0^\infty \frac{t}{4} dt 
\int_{\bbar\la=\tl}
\vev{\la~d\la} \cb{\tl~d\tl}~\delta(K^2 - t\gb{\la|K|\tl})
( \bullet ). 
\eea
This second delta function sets $t$ to the value
\bea
t= \frac{K^2}{\gb{\la|K|\tl}}
\Label{t-eqn}
\eea
so, taking account of the prefactor $\gb{\la|K|\tl}$ of $t$ inside the delta function, we can now perform the $t$-integral trivially:
\bea
 \int d\mu~~ (\bullet )  =
- \int_{\bbar\la=\tl}
\vev{\la~d\la} \cb{\tl~d\tl}~
 \frac{K^2}{4\gb{\la|K|\tl}^2} 
( \bullet ). 
\Label{spin-int-me}
\eea  
The remaining integral over the spinor variables is carried out with the residue theorem.  We will see how this is done in the master integrals before proceeding to the case of general amplitudes.

\subsubsection{Cut bubble}
Let us start with the scalar bubble integral.  The integrand consists entirely of the two cut propagators, so the cut is simply the integral of the LIPS measure,
\bea
\Delta\left( \frac{1}{\ell^2 (\ell-K)^2}\right)
= \int d\mu = -\int_{\bbar\la=\tl}
\vev{\la~d\la} \cb{\tl~d\tl}~
 \frac{K^2}{4\gb{\la|K|\tl}^2} .
\Label{cut-bub-spin}
\eea
In calculating coefficients of complete amplitudes, it can suffice to leave the cut bubble in the form (\ref{cut-bub-spin}) and work at the integrand level.  Here we continue and show how to apply the residue theorem to complete the integral.
We make use of the following identity to rewrite the integrand as a total derivative \cite{Cachazo:2004kj}.  Here $\eta$ is an arbitrary spinor.
\bea
[\W \la~d\W \la]
{1\over \gb{\la|K|\W \la}^2}=[d\W \la~\partial_{\W \la}]
\left( {[\W \la~\eta] \over  \gb{\la|K|\eta} \gb{\la|K|\lt}}\right).
\Label{tot-deriv}
\eea
However, the integral is not identically zero, because there are delta-function contributions along the contour.  In the theory of a complex variable, we know that
\bea
 {\partial \over \partial \bbar
z} {1\over (z-b)}=2\pi \delta(z-b).
\Label{hol-anom}
\eea
Therefore, we pick up a residue at the pole $\ket{\la} =|K|\eta]$.  Along the contour, since $\la$ and $\tl$ are conjugates, we also substitute $|\W \la]=|K|\eta\rangle$.  The result for the four-dimensional cut bubble  is thus
\bea
\Delta I_2  &=& {i \over \pi^2} \Delta\left( \frac{1}{\ell^2 (\ell-K)^2}\right)
=  - {i K^2 \over 2 \pi}
\left. \left( {[\W \la~\eta] \over   \gb{\la|K|\W \la}}
\right)\right|_{\ket{\la}=|K|\eta]}
= {1\over 2\pi i}.
\eea
(Different conventions in the literature such as in \cite{Britto:2005ha} and \cite{Mastrolia:2009dr} yield results with different  powers of $i$ and $2\pi$; again these will be unimportant as long as the framework is consistent.  The main point is clear, namely that the result is a constant, proportional to the discontinuity across the branch cut of the logarithmic function (\ref{masterbubble}). )

\subsubsection{Cut triangle}
In the unitarity cut of the scalar triangle, there is one propagator left over along with the LIPS measure.  Converting to the spinor variables, this factor is $(\ell+K_3)^2=t\gb{\la|K_3|\tl}+K_3^2$.  Performing the $t$ integral as before, making the substitution (\ref{t-eqn}) throughout, we have
\bea
\hspace{-0.5in}
\Delta\left( \frac{1}{\ell^2 (\ell-K)^2 (\ell+K_3)^2}\right)
= - \int_{\bbar\la=\tl}
\vev{\la~d\la} \cb{\tl~d\tl}~
 \frac{1}{4\gb{\la|K|\tl} \gb{\la|Q|\tl}},
\Label{cut-tri-spin}
\eea
where
\bea
Q=\frac{K_3^2}{K^2}K+K_3 .
\eea
Again, it is worth leaving the expression in the form (\ref{cut-tri-spin}), but let us see how to finish the integral.  The two factors in the denominator can be combined with a Feynman parameter, and the spinor integral done just as in the bubble case, so that we have
\bea
\hspace{-1in}
 - \int_0^1 dx  \int_{\bbar\la=\tl}
\vev{\la~d\la} \cb{\tl~d\tl}~
 \frac{1}{4\gb{\la|(1-x)K+xQ|\tl}}
= \frac{\pi}{2} \int_0^1 dx \frac{1}{((1-x)K+xQ)^2}.
\eea
The result for the cut in the $K$-channel is
\bea
\Delta I_3  &=& -{i \over \pi^2} \Delta\left( \frac{1}{\ell^2 (\ell-K)^2 (\ell-K_3)^2}\right)
\nonumber \\  &=& 
  {1 \over 2\pi i \sqrt{-\Delta_3}} 
\ln \left( \frac{-2(K_3^2+K\cdot K_3) +\sqrt{-\Delta_3}}{-2(K_3^2+K\cdot K_3) -\sqrt{-\Delta_3}}
\right),
\eea
where $\Delta_3$ is defined in (\ref{tri-kallen}).  Notice that this result is logarithmic, as expected.  Moreover, it is clear that all three-mass triangles are uniquely identified by the functions $\Delta_3$, which play a distinguished role in the expression as the arguments of square roots.  (For one-mass and two-mass triangles, $\Delta_3$ is a perfect square, so the square roots disappear from the formula while the logarithm remains.)

\subsubsection{Cut box}
The calculation of the cut scalar box integral is similar. Now there are two uncut propagators identifying the box, which we write as  $(\ell-K_i)^2$ and $(\ell-K_j)^2$.
 Converting to the spinor variables, they become  $K_i^2-t\gb{\la|K_i|\tl}$ and $K_j^2-t\gb{\la|K_j|\tl}$, respectively.  Performing the $t$ integral and making the substitution (\ref{t-eqn}) throughout, we have
\bea
\hspace{-1in}
\Delta\left( \frac{1}{\ell^2 (\ell-K)^2 (\ell-K_i)^2 (\ell-K_j)^2}\right)
= - \int_{\bbar\la=\tl}
\vev{\la~d\la} \cb{\tl~d\tl}~
 \frac{1}{4 K^2 \gb{\la|Q_i|\tl} \gb{\la|Q_j|\tl}},
\Label{cut-box-spin}
\eea
where now we define $Q_i, Q_j$ by
\bea
Q_i \equiv \frac{K_i^2}{K^2}K - K_i, \qquad
Q_j \equiv \frac{K_j^2}{K^2}K - K_j.
\eea
Here again, we can evaluate the integral by introducing a Feynman parameter.  It takes a form similar to the triangle.  The final result is
\bea
\Delta I_4  &=&  {1\over (2\pi i)2K^2 \sqrt{\Delta_{ij}}}\ln
\left({Q_i \cdot Q_j +\sqrt{\Delta_{ij}}\over Q_i \cdot Q_j -\sqrt{\Delta_{ij}}} \right),
\eea
where
\bea
\Delta_{ij} \equiv (Q_i \cdot Q_j)^2 - Q_i^2 Q_j^2.
\eea
We see that the cut is again logarithmic.
One can check that for either of the two choices of the cut configuration (straight across the box or selecting one corner), the function $\Delta_{ij}$ under the square root corresponds to the discriminant of the quadratic polynomial in (\ref{quadratic}).  In the cases where any of the corners of the box is a null momentum, $\Delta_{ij}$ is a perfect square.

\subsection{Unitarity cut of the amplitude}

To evaluate the cuts of master integrals, we needed only the relatively simple identity (\ref{tot-deriv}), from which we could find the residues according to (\ref{hol-anom}).  The cut of the full amplitude will generally have more complicated dependence on the loop momentum.  Let us examine its analytic structure.

\subsubsection{Stokes' Theorem on the complex plane}

Viewed formally, the residue theorem needed to evaluate cut integrals is an application of Stokes' Theorem to the complex plane \cite{Mastrolia:2009dr}.  Indeed, it is intuitively helpful to rewrite the cut integral in the familiar language of a single complex variable.  Given the cut momentum $K$, we can choose two null momenta $p$ and $q$ such that
\bea
K = p+q.
\eea
To see this, choose an arbitrary null vector $\W p$, and solve the equation $q = K- \alpha \W p $ such that $q^2=0$.  The solution is $\alpha = (\W p \cdot K)/(2 K^2)$.  Then set $p=\alpha \W p$.

The spinors $\la,\tl$ parametrizing loop momentum can now be expanded in the basis of spinors from $p$ and $q$
\bea
\lambda = \lambda_p + z \lambda_q, \qquad \tl = \tl_p + \bar z \tl _q.
\Label{zzbar}
\eea 
The homogeneity of the spinors has been used to set the coefficients of $\la_p$ and $\tl_p$ to 1, and now we see the familiar complex variables $z$  and $\bar z$ emerge as the other coefficients.  Recall that $\la,\tl$ were coordinates on $\IC\IP^1$. Equation (\ref{zzbar}) is a representation of the standard mapping of the complex plane to a sphere.

The cut integral measure now takes a reasonably simple form, since $\vev{\la d\la}[\tl d\tl]=-K^2 dz d\bar z$, and $\gb{\la|K|\tl}=K^2(1+z \bar z)$.  Thus the variable $t$ gets fixed to the value
\bea
t = {1 \over 1 + z \bar z},
\eea
and the cut measure (\ref{spin-int-me}) becomes
\bea
 \int d\mu~~ (\bullet )  =
 \int_{z^*=\bar z}
 \frac{dz d\bar z}{4(1+z \bar z)^2} 
( \bullet ). 
\eea

The cut integration can be performed by the Generalized Cauchy Formula.  
Given an integrand $F(z,\bar{z})$, construct a primitive $G(z,\bar{z})$ with respect to $\bar{z}$. Let $D$ be a disk in the complex plane containing all poles in $z$ of  $G(z,\bar{z})$. Then
\bea
\int_D   F(z,\bar{z}) ~d\bar{z} \wedge dz
=\oint_{\partial D} dz~ G(z,\bar{z}) 
- 2 \pi i \sum_{{\rm poles}~ z_j } {\rm Res} \{G(z,\bar{z}),z_j \}. 
\Label{gcf}
\eea

The bubble contribution is distinguished by being rational.

\subsubsection{Partial fraction identities for splitting denominator factors}

To look a little deeper into the structure of cuts of amplitudes, we return to the spinor variables $\la,\tl$.  Suppose we take the expressions for the tree amplitudes in the cut integral (\ref{cutdef}) from the Feynman rules.  The integrand is a rational function whose denominator is a product of propagator factors of the form $(\ell-K_i)^2$.  As we have seen the cuts of master integrals, such a factor becomes
\bea
(\ell-K_i)^2 = { K^2 \gb{\la|Q_i|\tl} \over \gb{\la|K|\tl} }.
\eea
Other factors of $\gb{\la|K|\tl}$ arise from the integral measure and the substitution for $t$ found in (\ref{t-eqn}).
The key property is that the denominator of the integrand consists of factors of $\gb{\la|Q_i|\tl}$, where no two $Q_i$ are the same, along with some power of the factor $\gb{\la|K|\tl}$.  

We find it helpful to rearrange the integrand in order to identify the cuts of  master integrals as given in (\ref{cut-bub-spin}), (\ref{cut-tri-spin}), and (\ref{cut-box-spin}).  This task is accomplished by partial fraction identities that split the denominator factors and reduce the power of 
$\gb{\la|K|\tl}$ if necessary.  In effect, it is a reduction technique for the cut integrals.

The splitting of factors with partial fractions proceeds as follows.  
First, split the factors $\gb{\ell|Q_j|\ell}$ among themselves, with the following identity: 
\bea
 {\prod_{j=1}^{k-1}[a_j~\tl]\over \prod_{i=1}^k
\gb{\la|Q_i|\tl}}=\sum_{i=1}^k {1\over
\gb{\la|Q_i|\tl}}{\prod_{j=1}^{k-1} \tgb{a_j|Q_i|\tl}\over
\prod_{m=1,m\neq i}^k \vev{\la|Q_m
Q_i|\tl}}.
\eea

Next,  reduce the power of $ \gb{\ell|K|\ell}$ in the remaining
denominators, since the master cuts contain at most one:
\bea 
\hspace{-0.5in}
{\prod_{j=1}^{n-1} [a_j~\tl]\over \gb{\la|K|\tl}^{n}
\gb{\la|Q|\tl}}  &=&  {\prod_{j=1}^{n-1}\tgb{a_j|Q|\tl}\over
\vev{\la|KQ|\la}^{n-1}} {1\over \gb{\la|K|\tl} \gb{\la|Q|\tl}}
\\ & & 
-
\sum_{p=0}^{n-2}
{
\left(
\prod_{j=1}^{n-p-2}\tgb{a_j|Q|\la}
\right)
\tgb{a_{n-p-1}|K|\la}
\left(
\prod_{t=n-p}^{n-1}[a_t~\tl]
\right)
\over
\gb{\la|K|\tl}^{p+2}
\vev{\la|K Q|\la}^{n-p-1}}.
\nonumber
\eea
Power-counting arguments ensure that  enough appearances of $\tl$ in the numerator to implement these identities as often as necessary.

It remains to implement the residue theorem, with the help of a generalized version of the differentiation identity (\ref{tot-deriv}) and the careful treatment of higher-multiplicity poles arising in the factor $\vev{\la|KQ|\la}$.

The procedure can be performed in generality.  Formulas for the coefficients are given below.

\subsection{Solutions for coefficients}

We now list formulas for the coefficients of master integrals \cite{Britto:2006fc,Britto:2007tt,Britto:2008sw,Britto:2008vq,Feng:2008ju}, through $\ord(\eps)$.  These references include $D$-dimensional versions of these formulas for going to higher orders in $\eps$, and with possible scalar masses.  These generalizations will be described in sections to follow, but the more general formulas will not be reproduced here. 

Our starting point is a general form for a single term (i.e., the numerator is a monomial in the loop momentum) in the sum of the {\em cuts} of one-loop Feynman integrals,
\bea
 i (4\pi)^{\frac{D}{2}} 
\int \frac{d^{D}\ell}{(2\pi)^{D}}  \delta^{(+)}(\ell^2)~\delta^{(+)}((K-\ell)^2){\cal T}^{(N)}(\ell).
\eea
Here $N$ is defined as the degree of ${\cal T}^{(N)}(\ell)$: it is the degree of the numerator minus the degree of the denominator, after setting $\ell^2$ to zero everywhere it appears. In renormalizable gauge, $N \leq 2$.
For convenience we also define some notation for the propagators,
\bea
Q_i &\equiv& \frac{K_i^2}{K^2}K - K_i. \\
D_i(\ell) &\equiv& (\ell-K_i^2) = K_i^2 - K^2 \frac{\gb{\la|K_i|\tl}}{\gb{\la|K|\tl}} = \frac{K^2 \gb{\la|Q_i|\tl}}{\gb{\la|K|\tl}}.
\eea
where the last equations are valid within the cut integral.
Finally, the cut integral should be recast in terms of the spinors $\la,\tl$ with the replacement 
\bean
\ell=t\la\tl, 
\eean
and the value of $t$ taken from (\ref{t-eqn}), 
\bean
t= \frac{K^2}{\gb{\la|K|\tl}}.
\eean

\subsubsection{Box coefficients} 

The coefficient of the box identified by the two cut propagators along with $D_r$ and $D_s$ is given by
\bea
\hspace{-0.5in}
  C[K_r,K_s,K]  =  {1\over 2} 
\left.\left(
{\cal T}^{(N)}(\ell) D_r(\ell) D_s(\ell)
\right)\right|_{\la \to P_{sr,1},\tl \to P_{sr,2}}
+ \{P_{sr,1}\leftrightarrow P_{sr,2}\}
\Label{box-formula}
\eea
Here we use the following definitions.  The vectors $P_{sr,1}$ and $P_{sr,2}$ are the null linear combinations of $Q_r$ and $Q_s$.
\bea
P_{sr,1} &=& Q_s + \left( {-Q_s \cdot Q_r + \sqrt{\Delta_{sr}}\over Q_r^2} \right) Q_r, \\
P_{sr,2} &=& Q_s + \left( {-Q_s \cdot Q_r - \sqrt{\Delta_{sr}}\over Q_r^2} \right) Q_r, \\
\Delta_{sr} &=& (Q_s \cdot Q_r)^2- Q_s^2 Q_r^2. 
\eea

\subsubsection{Triangle coefficients} 

If $N <-1$, the triangle coefficients are zero.  If $N \geq -1$, the coefficient of the triangle identified by the two cut propagators along with  $D_s$ is given by
\bea
\hspace{-0.5in}
C [K_s,K]  &=&
  {1\over 2(N+1)!\sqrt{\Delta_s}^{N+1}
\vev{P_{s,1}~P_{s,2}}^{N+1}} 
\\ & & \times \frac{d^{N+1}}{d\tau^{N+1}}
\left.\left(\left.
{\cal T}^{(N)}(\ell) D_s(\ell) \gb{\la|K|\tl}^{N+1}
\right|_{\tl \to Q_s \la ,\la \to P_{s,1}- \tau P_{s,2}}
\right.\right.
\nonumber \\ & &
~~~~~~~~~~~~~~~~+ \{P_{s,1}\leftrightarrow P_{s,2}\}
\Bigg)
\Bigg|_{\tau \to 0}
\nonumber
\eea
Here we use the following definitions. The vectors $P_{s,1}$ and $P_{s,2}$ are null linear combinations of $Q_s$ and $K$.
\bea
P_{s,1} &=& Q_s + \left({-Q_s \cdot K + \sqrt{\Delta_{s}}\over K^2} \right) K,
 \\
P_{s,2} &=& Q_s + \left({-Q_s \cdot K - \sqrt{\Delta_{s}}\over K^2} \right) K,
\\ \Delta_{s} &=& (Q_s \cdot K)^2- Q_s^2 K^2.
\eea
The effect of the multiple derivative in the parameter $\tau$, evaluated at $\tau=0$, is simply to pick out a term in the series expansion.

\subsubsection{Bubble coefficient} 

There is just one bubble in the cut channel $K$.  If $N<0$, the coefficient is zero.  If $N \geq 0$, the coefficient is
\bea
\hspace{-1in}
C[K] = K^2  \sum_{q=0}^N {(-1)^q\over q!} {d^q \over
ds^q}\left.\left( {\cal B}_{N,N-q}^{(0)}(s)
  + \sum_{r=1}^k\sum_{a=q}^N
\left({\cal B}_{N,N-a}^{(r;a-q;1)}(s)-{\cal
B}_{N,N-a}^{(r;a-q;2)}(s)\right)\right)\right|_{s=0},
\eea
where
\bean 
\hspace{-1in}
 {\cal B}_{N,m}^{(0)}(s)\equiv {d^N\over d\tau^N}\left.\left(
\left.
{(2\eta\cdot
K)^{m+1}  \gb {\la|K|\tl}^N  \over N! [\eta|\eta' K|\eta]^{N}(m+1) (K^2)^{m+1} \vev{\la~\eta}^{N+1} }
{\cal T}^{(N)}(\ell)
\right|_{\stackrel{\tl \to (K+s \eta)\cdot\la}{\la \to (K-\tau \eta')\cdot\eta}}
\right)
\right|_{\tau \to 0},
\eean
\bean
\hspace{-1in}
 {\cal B}_{n,m}^{(r;b;1)}(s)  \equiv  & & {(-1)^{b+1}\over
 b! (m+1) \sqrt{\Delta_r}^{b+1} \vev{P_{r,1}~P_{r,2}}^b} \times
\\ & & 
{d^b \over d\tau^{b}}
\left.
\left( {\gb{\la|\eta|P_{r,1}}^{m+1} 
\vev{\la|Q_r \eta|\la}^{b} \gb {\la|K|\tl}^{N+1} 
\over
\gb{\la|K|P_{r,1}}^{m+1} \vev{\la|\eta K|\la}^{n+1} }
{\cal T}^{(N)}(\ell) D_r(\ell)
\right)
\right|_{\stackrel{\tl \to (K+s \eta)\la,~\la \to P_{r,1}-\tau P_{r,2}}{\tau=0}} 
\eean
\bean
\hspace{-1in}
 {\cal B}_{n,m}^{(r;b;2)}(s)  \equiv  & & {(-1)^{b+1}\over
 b! (m+1) \sqrt{\Delta_r}^{b+1} \vev{P_{r,1}~P_{r,2}}^b} \times
\\ & & 
{d^b \over d\tau^{b}}
\left.
\left( {\gb{\la|\eta|P_{r,2}}^{m+1} 
\vev{\la|Q_r \eta|\la}^{b} \gb {\la|K|\tl}^{N+1} 
\over
\gb{\la|K|P_{r,2}}^{m+1} \vev{\la|\eta K|\la}^{n+1} }
{\cal T}^{(N)}(\ell) D_r(\ell)
\right)
\right|_{\stackrel{\tl \to (K+s \eta)\la,~\la \to P_{r,2}-\tau P_{r,1}}{\tau=0}} 
\eean
Here $\eta,\eta'$ are arbitrary spinors; they should be generic in the sense that they do not coincide with any spinors from massless external legs.

\section{Generalized unitarity: unphysical cut channels}

Although one-loop amplitudes are constructible by their branch cuts in physical momentum channels, and we have seen how this construction can be carried out, it can be useful to solve for the coefficients of master integrals by studying other singularities.

 Unitarity cuts can be ``generalized'' in the sense of putting a different number of propagators on shell.  This operation selects different kinds of singularities of the amplitude; they are not physical momentum channels like ordinary cuts and do not have an interpretation relating to the unitarity of the S-matrix.  Here, it becomes essential to work with complexified momenta.

\subsection{Quadruple cuts}

\begin{figure}
\begin{center}
\includegraphics[width=8cm]{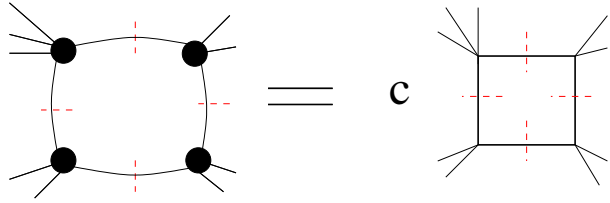}
\end{center}
\caption{A quadruple cut puts four propagators on shell.  It is a trivial integral isolating a single box coefficient.}
\label{fig:quadcut}
\end{figure}

The most direct application of generalized unitarity is to use a ``quadruple cut'' to find any box coefficient \cite{Britto:2004nc}.
If we cut {\it four} propagators---equivalent to specifying a partition $(K_1,K_2,K_3,K_4)$ of the external momenta---then the four-dimensional integral becomes trivial.  See Figure \ref{fig:quadcut}. 
\bea
\Delta_4 A^{\rm 1-loop} = \int d^4\ell~ \delta(\ell_1^2) ~\delta(\ell_2^2) ~\delta(\ell_3^2) ~\delta(\ell_4^2)~ A_1^{\rm tree} A_2^{\rm tree} A_3^{\rm tree} A_4^{\rm tree} 
\eea
 Applied to the master integrals, the quadruple cut picks up a contribution from exactly one box integral, namely the one with momenta $(K_1,K_2,K_3,K_4)$ at the corners.  Therefore, the cut expansion collapses to a single term:
\bea
\Delta_4 A^{\rm 1-loop} = c_4(K_1,K_2,K_3,K_4) \Delta_4 I_4(K_1,K_2,K_3,K_4).
\eea
The quadruple cut of the scalar box integral is  a Jacobian factor which is equal on both sides of the equation.  The result for the coefficient is simply
\bea
c_4 = 
 \frac{1}{2} \sum_{\ell \in \cal S} A^{\rm tree}_1(\ell) A^{\rm tree}_2(\ell) A^{\rm tree}_3(\ell) A^{\rm tree}_4(\ell),
\label{qcutboxcoeff}
\eea
where ${\cal S}$ is the solution set for the four delta functions of the cut propagators,
\bea
\hspace{-0.5in}
{\cal S}=\{\ell~ | \ell^2=0, \quad (\ell-K_1)^2=0, \quad (\ell
-K_1-K_2)^2=0, \quad (\ell+K_4)^2=0\}.
\eea
There are exactly two solutions, provided that momenta are allowed to take complex values.  This is the origin of the factor of 2 in the denominator of (\ref{qcutboxcoeff}).
Thus it is easy to get all the box coefficients. 

In effect, the substitutions in the formula given previously in equation (\ref{box-formula}) implement the quadruple cut solutions directly.

\subsection{Triple cuts}

  A triple cut, in which three selected propagators are put on shell, targets the coefficient of the unique scalar triangle including those three propagators.  However, it also picks up contributions from all scalar boxes with those same three propagators.  Moreover, there is a one-dimensional integral left over, so the calculation is less direct than the quadruple cut.

Forde's parametrization of loop momentum allows the extraction of triangle coefficients from triple cuts \cite{Forde:2007mi}, for the case of massless propagators.  Suppose the triple-cut conditions are given by
\bea
\ell^2=0, \qquad (\ell-K_1)^2 = 0, \qquad (\ell+K_3)^2=0.
\Label{triple-cut}
\eea
From $K_1$ and $K_3$, construct null vectors as follows:
\bea
K_1^\flat = \gamma \alpha \frac{\gamma K_1 -S_1 K_3}{\gamma^2-S_1 S_3},
\qquad
K_3^\flat = \gamma \alpha' \frac{\gamma K_3-S_3 K_1}{\gamma^2-S_1 S_3},
\eea
where $S_1=K_1^2$, $S_3=K_3^2$, and 
\bea
\hspace{-1in}
\gamma=K_1 \cdot K_3 \pm \sqrt{(K_1 \cdot K_3)^2 - K_1^2 K_3^2},
\quad
\alpha = \frac{S_3(S_1-\gamma)}{S_1 S_3 - \gamma^2},
\quad
\alpha' = \frac{S_1(S_3-\gamma)}{S_1 S_3 - \gamma^2}.
\eea 
Then the loop momentum can be expressed in terms of a single parameter $t$ such that it satisfies the three cut conditions (\ref{triple-cut}) explicitly.  The parametrization is 
\bea
\ell = K_1^\flat + K_3^\flat + {t \over 2}\vev{K_1^{\flat,-}|\gamma^\mu|K_3^{\flat,-}} + {1 \over 2t}\vev{K_3^{\flat,-}|\gamma^\mu|K_1^{\flat,-}} .
\eea
Plugging this expression into the integrand executes the triple cut, giving us the triangle  of interest along with some boxes.  Each box contribution has an additional propagator, with two poles in $t$.  Therefore the box terms can be removed by expanding the expression around the point $t = \infty$.
The result for the triangle coefficient is given by
\bea
- \left. [{\rm Inf}_t A_1 A_2 A_3](t) \right|_{t=0} ,
\eea
averaged over the two solutions for $\gamma$,
where $A_1,A_2,A_3$ are the three tree-level contributions analogous to the amplitudes in (\ref{cutdef}), and the function ${\rm Inf}$ denotes the series expansion around $t = \infty$.

Another approach to triple cuts \cite{Mastrolia:2006ki} is to write the third delta function as the difference of two propagators with different $i\varepsilon$ prescriptions, 
\bea
2\pi i \delta(p^2) \to \frac{1}{p^2 + i \varepsilon} - 
\frac{1}{p^2 - i \varepsilon},
\eea
and treat the problem as a double cut in the first two delta functions with modified propagators appearing in the tree-level amplitudes.  Propagator modification is also the key to Feynman's Tree Theorem, relating loop amplitudes to sums of products of tree amplitudes produced by on-shell delta function insertions.  The concept fits in the framework of generalized unitarity, so we will describe it briefly in the following subsection.

\subsection{Feynman's Tree Theorem}

Feynman observed \cite{FeynmanTT-1,FeynmanTT-2} that any (Feynman) diagram with closed loops can be expressed in terms of tree diagrams, in the same spirit as Cutkosky but with more general cuts.  It is not exactly the type of ``generalized unitarity'' described above for triple and quadruple cuts, because the propagators have been changed.  We illustrate it in the simplest case, with scalar propagators at one loop.

The key identity relates Feynman propagators $G_F$ and advanced propagators $G_A$, which differ by a delta function.  (Similarly, one could choose to use retarded instead of advanced propagators throughout.)  Let $k=(k_0,\vec{k})$ be the 4-momentum of the scalar,  $m$ its mass, and $\omega=\sqrt{|\vec{k}|^2+m^2}$.  Then
\bea
G_F(k)
&= \frac{i}{2 \omega} \left[
\frac{1}{k_0-\omega+i\varepsilon}-\frac{1}{k_0+\omega-i\varepsilon}
\right]
 &= \frac{i}{k_0^2 - \omega^2 + i\varepsilon} 
\\
G_A(k) 
&= \frac{i}{2 \omega} \left[
\frac{1}{k_0-\omega-i\varepsilon}-\frac{1}{k_0+\omega-i\varepsilon}
\right]
&= \frac{i}{k_0^2 - \omega^2 - i\varepsilon ~{\rm sgn}(k_0)} 
\eea
Using the identity
\bea
\frac{1}{x \pm i\varepsilon} = PV\left(\frac{1}{x}\right) \mp i \pi \delta(x),
\eea
where PV denotes the principal value prescription, we can see that
\bea
G_A(k) = G_F(k) - 2 \pi  \delta^{(+)}(k^2-m^2).
\Label{id-af}
\eea
Now consider the loop integral in which every propagator is replaced by an advanced propagator.  All the poles are located above the real axis of $\omega$, so we can close the contour below, and the result is zero.  But when we apply the identity (\ref{id-af}), we get
\bea
0 = \int_k N(k) \prod_i G_A^{(i)}(k) 
= \int_k N(k) \prod_i \left( G_F(k) - 2 \pi  \delta^{(+)}(k^2-m^2) \right).
\eea 
Here $N(k)$ denotes the numerator.  Expanding the product on the right-hand side, the first term is the original physical loop integral, and every other term has at least one on-shell delta function.  Feynman's Tree Theorem is this expression of the physical loop integral in terms of tree diagrams derived from its on-shell cuts.

Recently, this theorem has been been used in the proof of covariance of the MHV prescription at one loop \cite{Brandhuber:2005kd}.  It is  interesting to try to compute amplitudes directly from the tree-level ingredients with altered propagator prescriptions \cite{CaronHuot:2010zt,Catani:2008xa,Bierenbaum:2010cy}.  Compared to other ``unitarity'' methods, those based on Feynman's Tree Theorem work without reference to master integrals, as the entire integral is contained in the proper description of the propagators and their analytic continuations.

\subsection{Single cuts}

Quadruple cuts compute box coefficients.  Triple cuts compute triangle coefficients, and box coefficients as well.  Traditional double cuts compute bubble, triangle, and box coefficients, as we saw in the previous section.  Single cuts should then give all the coefficients of master integrals.  

It is not easy to compute single cuts analytically. They are relatively unphysical and more strongly divergent than double (or triple or quadruple) cuts.
When a single propagator is cut, the  integrand is typically a tree-level amplitude with singular kinematics.  Exceptionally, the kinematics are nonsingular for massless particles in supersymmetric gauge theories and some massive particles with additional supersymmetry \cite{CaronHuot:2010zt}, and they have been used to study higher-loop integrands in $\N=4$ SYM \cite{ArkaniHamed:2010kv}.

Nevertheless, some QCD amplitudes have been computed from single cuts alone \cite{NigelGlover:2008ur}; single cut integrals have been analyzed and a prescription for their evaluation formulated in analogy with the residue theorem for double cuts \cite{Britto:2010um}; and a single-cut method has been found using the principles of Feynman's Tree Theorem but grouping the terms together into a smaller set of single-cut amplitudes \cite{Catani:2008xa}.

\section{$D$-dimensional unitarity\label{sec:ddim}}

We have seen how to evaluate four-dimensional cuts to solve for the scalar box, triangle, and bubble coefficients in the expansion of a one-loop amplitude.  In a four-dimensional expansion, there are additional cut-free ``rational'' terms.  Traditional reduction techniques start with a loop integral and reduce it to a set of master integrals using linear relations generating the rational coefficients.  None of the master integrals is purely rational.  Where, then, do the purely rational terms come from?  The answer is that they arise in the expansion in $\eps$.  Higher-order corrections to the coefficients combine with the UV-divergent parts of the integrals to produce a purely rational term at $\ord(\eps^0)$.  
\bea
A^{\rm 1-loop} = \sum_n \sum_{{\bf K}=\{K_1,\ldots,K_n\}} 
\left. c_n({\bf K}) \right|_{\eps=0}
 I_n({\bf K}) + {\cal R} + \ord(\eps)
\Label{basis-exp-rational}
\eea
If we keep full dependence in $\eps$ without the expansion, there are no terms apart from the master integrals in the original expansion 
\bea
A^{\rm 1-loop} = \sum_n \sum_{{\bf K}=\{K_1,\ldots,K_n\}} 
c_n({\bf K})(\eps) \times
 I_n({\bf K}).
\eea
  This property is the basis of $D$-dimensional cut-constructibility \cite{vanNeerven:1985xr,Bern:1992em,Bern:1993kr,Bern:1995db}.  We need to generalize our cut integral to $D$ dimensions to pick up the higher-order corrections to the coefficients.  Here we present the physical double-cut integral generalizing the treatment in Section \ref{sec:umethod} \cite{Anastasiou:2006gt,Anastasiou:2006jv}.  A similar analysis for generalized cuts is available as well \cite{Badger:2008cm}.

All external momenta remain four-dimensional.
We decompose the loop momentum $\ell$  into its four-dimensional part $\W\ell$, and the remaining $\mu$ which is orthogonal \cite{'tHooft:1972fi,Mahlon:1993fe,Mahlon:1993si, Bern:1995ix, Bern:1995db}.  Then the integration measure factorizes, 
\bea 
& &
\int d^{4-2\eps} \ell  =  \int {d^{-2\eps} \mu } \int {d^{4}
\W\ell}  =
{ { (4\pi)^{\eps}\over \Gamma(-\eps)}} \int_0^1  du~
u^{-1-\eps}\int {d^{4} \W\ell}.
\eea
  We have introduced the dimensionless variable $u$,
defined by 
\bea
u = {4 \mu^2 \over K^2}.
\eea

In the $D$-dimensional unitarity approach, we proceed by working on the familiar four-dimensional integral, which now depends on the new variable $u$.  We keep the $u$-dependence throughout.  Dropping the prefactor ${ (4\pi)^{\eps} / \Gamma(-\eps)}$, the cut measure is then
\bea \int_0^1 du~ u^{-1-\eps} \int d^4\W\ell ~
\delta^{(+)}(\W\ell^2-\mu^2) \delta^{(+)}((\W\ell-K)^2-\mu^2)
\eea
To make use of the spinor technology that is so effective for four-dimensional cuts, we want to relate $\W\ell$ to a null 4-vector.  This can be done with the help of the cut vector $K$.  We let
\bea
\W\ell = p + z K
\eea
where $p$ should be null.  This condition is implemented by a delta function on $p^2$ together with an integral over $z$.
\bea \int_0^1 du~ u^{-1-\eps} \int dz \int d^4\W\ell ~
\delta^{(+)}(\W\ell^2-\mu^2) \delta^{(+)}((\W\ell-K)^2-\mu^2) \delta(p^2)
\eea
Now we have three delta functions.  After some manipulation of their arguments, we are able to perform the $z$-integral trivially, finding that 
\bea
z = \frac{1-\sqrt{1-u}}{2}.
\eea
The remaining cut measure is 
\bea
\int_0^1 du~ u^{-1-\epsilon}
\int d^4 p~~ \delta(p^2) ~\delta(\sqrt{1-u}~ K^2-2 K \cdot p).
\Label{u-int}
\eea
At this point we can replace the null momentum with spinor variables as before, with $p=t\la\tl$.  The cuts of the master integrals are modified as follows from the four-dimensional versions in Section \ref{sec:umethod}.
\bea
\hspace{-1in} 
\Delta I_2= {1 \over 2\pi i}\int_0^1 du~u^{-1-\eps}  \sqrt{1-u}
\eea

\bea
\hspace{-1in} 
\Delta I_3   =
  -{1 \over 2\pi i } \int_0^1 du~u^{-1-\eps}
{1\over \sqrt{-\Delta_{3}}}
\ln \left( {Z +\sqrt{1-u}\over Z-\sqrt{1-u}
}\right)
,
\eea
with
\bea Z=-{K\cdot K_3+ K_3^2 \over \sqrt{ (K\cdot K_3)^2-
K^2 K_3^2}}.~~~\label{I3m-para}\eea
\bea
\hspace{-1in}
\Delta I_4  =  \int_0^1 du~u^{-1-\eps}{1\over (2\pi i)2K^2 \sqrt{\Delta_{ij}-A u}}\ln
\left({Q_i \cdot Q_j - C u +\sqrt{1-u}\sqrt{\Delta_{ij}-A u}\over Q_i \cdot Q_j -C u -\sqrt{1-u}\sqrt{\Delta_{ij}-A u}} \right),
\eea
where
\bea
  & A = -{1\over K^2}\det\left( \begin{array}{ccc} K_i^2~~ & K_i\cdot
K_j~~ & K_i\cdot K
\\ K_i\cdot K_j~~ & K_j^2~~ & K_j\cdot K \\
K_i\cdot K~~ & K_j\cdot K~~ &
K^2\end{array}\right),~~~~~
& B = \Delta_{ij},
\\
 & C = {1\over K^2}\det\left(
\begin{array}{cl}  K_i\cdot K_j~~ & K_i\cdot K \\ K_j\cdot K~~ & K^2
\end{array}\right),~~~~~
&D = Q_i \cdot Q_j.
\eea
Finally, we add the scalar pentagon to the basis of master integrals.  
Here we have three momenta $K_i,K_j,K_k$ appearing in the three uncut propagators, and we define $Q_i,Q_j,Q_k$ by
\bea
 Q_r & = &  - (\sqrt{1-u})K_r+ \frac{K_r^2- (1-\sqrt{1-u})( K_r\cdot K)}{K^2} K,
\eea
The cut is given by
\bea
\Delta I_5 &=&
 {1 \over 2 \pi i} \int_0^1 du~ u^{-1-\eps}
 {\sqrt{1-u} \over (K^2)^2}
~~~\label{Pentagon-gen} \\ &  &
\left({ S[Q_3, Q_2, Q_1,K]\over
4\sqrt{(Q_3\cdot Q_2)^2 -Q_3^2 Q_2^2}} \ln{  Q_3 \cdot Q_2 -
\sqrt{  (Q_3\cdot Q_2)^2 -Q_3^2 Q_2^2}\over  Q_3 \cdot Q_2 +
\sqrt{  (Q_3\cdot Q_2)^2 -Q_3^2 Q_2^2}}\right. 
\nonumber \\ & & + { S[Q_3,
Q_1, Q_2,K]\over 4\sqrt{(Q_3\cdot Q_1)^2 -Q_3^2 Q_1^2}} \ln{
Q_3 \cdot Q_1 - \sqrt{  (Q_3\cdot Q_1)^2 -Q_3^2 Q_1^2}\over  Q_3
\cdot Q_1 + \sqrt{  (Q_3\cdot Q_1)^2 -Q_3^2 Q_1^2}}  
\nonumber \\ & &
\left.+ { S[Q_2, Q_1, Q_3,K]\over 4\sqrt{(Q_2\cdot Q_1)^2
-Q_2^2 Q_1^2}} \ln{  Q_2 \cdot Q_1 - \sqrt{  (Q_2\cdot Q_1)^2
-Q_2^2 Q_1^2}\over  Q_2 \cdot Q_1 + \sqrt{ (Q_2\cdot Q_1)^2
-Q_2^2 Q_1^2}}\right),
\nonumber
\eea
where
\bea
 S[Q_i, Q_j, Q_k,K] \equiv
 \frac{ 2 \det \left( \begin{array}{lcr} K \cdot Q_k & Q_i \cdot K & Q_j \cdot K\\
Q_i \cdot Q_k & Q_i^2 & Q_i \cdot Q_j \\ Q_j \cdot Q_k & Q_i \cdot Q_j &
Q_j^2\end{array} \right)}{ \det \left( \begin{array}{lcr} Q_k^2 & Q_i \cdot Q_k & Q_j \cdot Q_k\\
Q_i \cdot Q_k & Q_i^2 & Q_i \cdot Q_j \\ Q_j \cdot Q_k & Q_i \cdot Q_j &
Q_j^2
\end{array} \right)} .
~~~\label{Func-S-sec}
\eea

The full cut amplitude produces similar functional forms.  As in the four-dimensional case, the pieces coming from different master integrals are identifiable by their unique logarithms.  The pentagons are an exception: the arguments of their logarithms are the same as from the boxes.  The cut pentagons are distinguished from cut boxes just by the rational prefactor in the functional form $S[Q_i,Q_j,Q_k,K]$.  In our approach, we use the logarithmic function to extract the box and pentagon coefficients simultaneously as total ``box'' coefficients, and later separate the pentagon parts from the box parts.

The cut amplitude is naturally decomposed into linear combinations of the functions above, the cuts of master integrals, with coefficients rational in the momentum invariants---and now with additional dependence on the parameter $u$.  This dependence turns out to be purely polynomial \cite{Britto:2008sw}.  This behavior is expected on physical grounds but is not immediately apparent due to the irrational factor $\sqrt{1-u}$ now appearing in the formulas.

Because the $u$-dependence of the coefficients is polynomial, we do not need to finish the $u$ integral explicitly.  Instead, we use recursion/reduction formulas, derived from integration by parts, to deal with the higher powers of $u$.  More specifically, terms in the cut amplitude will emerge naturally as combinations of the following functions.
\begin{eqnarray}
&& {\rm Bub}^{(n)} =  \int_0^1 du u^{-1-\eps} u^n
\sqrt{1-u} \label{eq:bubmaster}
\\ \nn
&& {\rm Tri}^{(n)} = \int_0^1 du u^{-1-\eps} u^n\ln
\left( {Z +\sqrt{1-u}\over Z-\sqrt{1-u}
}\right) 
\label{eq:trimaster}
\\ \nn
&& {\rm Box}^{(n)} = \int_0^1 du u^{-1-\eps} {u^n\over
\sqrt{B - A u}}  \ln \left( {D - C u- \sqrt{1-u}\sqrt{
B - A u}\over D - C u+ \sqrt{1-u}\sqrt{
B - A u}}\right)
\label{eq:boxmaster}
\end{eqnarray}
These functions are related to the cut master integrals by
\begin{eqnarray}
&&\hspace{-0.5cm}  {\rm Bub}^{(n)} = F^{(n)}_{2 \to 2} {\rm Bub}^{(0)}
\nonumber \\
&& \hspace{-0.5cm} {\rm Tri}^{(n)}(Z) = F^{(n)}_{3 \to 3}(Z) {\rm Tri}^{(0)}(Z)
+ F^{(n)}_{3\to 2}(Z) {\rm Bub}^{(0)}
\nonumber \\
&&\hspace{-0.5cm} {\rm Box}^{(n)}
 = F^{(n)}_{4 \to 4} {\rm Box}^{(0)}
+ \bigg \{ 
F^{(n)}_{4 \to 3}(Z_1) {\rm Tri}^{(0)}(Z_1) 
+ F^{(n)}_{4\to 2}(Z_1) {\rm Bub}^{(0)}
+ (Z_1 \leftrightarrow Z_2) \bigg\}
\nonumber \\
&&\hspace{-0.5cm}  F^{(n)}_{2 \to 2} =  {(-\eps)_{3 \over 2} \over (n-\eps)_{3 \over 2}}, \quad F^{(n)}_{3 \to 3} =  {-\eps\over n-\eps} (1-Z^2)^n,
\nonumber \\
&& \hspace{-0.5cm}
 F^{(n)}_{4 \to 4} = { (-\eps)_{1 \over 2} \over (n-\eps)_{1 \over 2}}  
\left(  {B \over A} \right)^n,
\nonumber \\
&&\hspace{-0.5cm}  F^{(n)}_{3 \to 2} =  {(-\eps)_{3 \over 2} \over n-\eps}  \sum_{k=1}^n {{2 Z}(1-Z^2)^{n-k} \over (k-\eps)_{1 \over 2}}
\nonumber \\
&&\hspace{-0.5cm} F^{(n)}_{4 \to j} ={D + (Z^2-1) C \over (n-\eps)_{1 \over 2} Z A } \sum_{k=1}^n
\left(  {B \over A} \right)^{n-k} \hspace{-0.4cm}{F^{(k-1)}_{3 \to j}\over (k-1/2-\eps)_{1 \over 2}}
\label{eq:reduction}
\end{eqnarray}
Here $(x)_n = \Gamma(x+n)/\Gamma(x)$, and $j=2,3$. In the equation for the box shift, $Z_1,Z_2$ correspond to the 
two possible cut-triangles  obtained by pinching the uncut propagators of the box. 

Alternatively, the result can simply be given in terms of dimensionally shifted master integrals.

\section{Treatment of massive particles}

In the preceding sections, we have explored unitarity cuts of amplitudes whose internal lines are all massless.  When internal lines can come from massive particles, the formulas need to be modified.

Scalar masses are fairly straightforward to deal with after having studied $D$-dimensional cuts.  
In dimensional regularization, the quantity $\mu^2$ behaves as a scalar mass for the four-dimensional parts of the loop momenta.  It is not difficult to generalize the formulas for cuts and coefficients to account for different values of masses \cite{Britto:2006fc,Kilgore:2007qr,Britto:2008vq}.  We can immediately calculate the coefficients of bubbles, triangles, boxes and pentagons.

A bigger issue is that new types of master integrals arise:  the scalar tadpole, and ``massless bubbles,'' which have a single massless external leg on one side ($K^2=0$) and a massive particle circulating in the loop.  

The scalar tadpole integral is
\bea I_1 & \sim & m^{2-2\eps} { \Gamma[1+\eps]\over \eps
(\eps-1)} \\
& = & {m^2 \over \eps} + m^2 (1-\gamma-\log(m^2))+\ord(\eps),
~~~~\label{One-point} \eea
and the massless bubble integral is
\bea 
\hspace{-0.5in}
I_2(K=0) &\sim& {M_1^{2-2\eps}-M_2^{2-2\eps} \over M_1^2 - M_2^2
}
{ \Gamma[1+\eps]\over \eps (1-\eps)}\\
&=&
{1 \over \eps} + (1-\gamma)-{1 \over M_1^2-M_2^2}
(M_1^2\log(M_1^2)-M_2^2\log(M_2^2))+\ord(\eps)
\\ & {\rm or} &
{1 \over \eps} -(\gamma  + \log M_1^2)
+\ord(\eps)~~~~{\rm if}~~M_1=M_2.
\eea

In massless theories, both of these integrals are set to zero in dimensional regularization.  Neither of these integrals has a branch cut in any physical channel.  

Another problematic integral is a bubble with a single massive fermion on one side, which is also cut-free in four dimensions, proportional to $(m^2)^{-\eps}$.
However, the diagrams contributing to the master integral expansion contain integrals that do have physical cuts in the $m^2$ channel, as well as contributions from self-energy insertions on an external line.  One proposed resolution involves applying a unitary method in a specific gauge so that the wavefunction renormalization factor takes a fixed value \cite{Ellis:2008ir}.

The coefficients of the cut-free integrals can, in some cases, be completely fixed by the universal ultraviolet and infrared divergent behavior \cite{Giele:1991vf,Kunszt:1994mc,Bern:1995db, Mitov:2006xs, Badger:2008za}.

Within the unitarity method, there is a proposal to find massless bubble coefficients from taking a suitable limit and the tadpole coefficients from an artificial double cut, constructed by introducing an auxiliary, unphysical propagator into the formulas \cite{Britto:2009wz}, based on the integrand classification of \cite{Ossola:2006us}.  The artificial double cut can be evaluated by standard unitarity methods.

Finally, within generalized unitarity, techniques for computing single cuts analytically can be used to derive some QCD amplitudes \cite{NigelGlover:2008ur,Britto:2010um}.  Single cuts exhibit some pathological behavior in general and need to be treated carefully.  The subject deserves further study.

\section{MHV diagrams}

A discussion of new techniques for loop amplitudes would not be complete without mentioning the applications of MHV diagrams by Cachazo-Svr\v{c}ek-Witten (CSW) rules as a tool in amplitude construction.  

MHV diagrams and CSW rules \cite{Cachazo:2004kj} are an alternative to and on-shell analog of Feynman diagrams and rules.  The external legs are in fixed helicity states.  The vertices are MHV amplitudes, and there is a prescribed on-shell continuation for propagators to reconcile momentum conservation and intermediate states of definite helicity.  At tree level, it is understood that they are valid because they reproduce all the correct singularities.  

The MHV diagram technique can be applied in several ways at loop level.  Most immediately, it can be used to construct tree-level amplitudes for use in a (possibly generalized) unitarity cut integral, and the four-dimensional cuts evaluated as above.

Similarly, cut propagators that have been analytically continued to lie on shell can be continued further to $D=4-2\eps$ dimensions; one can then continue with explicit evaluation  of a $D$-dimensional cut integral, which was used in \cite{Brandhuber:2004yw,Quigley:2004pw,Bedford:2004py,Bedford:2004nh} to recover one-loop MHV amplitudes in Yang-Mills theory with various amounts of supersymmetry.
Extensions of CSW rules to include massive scalars \cite{Boels:2007pj} lead to MHV-diagrammatic constructions of rational terms in one-loop gluon amplitudes \cite{Boels:2008ef}.

The CSW rules were conjectured based on the localization of Yang-Mills amplitudes in twistor space.  Along these lines, twistor space actions were written that reproduce the MHV diagram construction \cite{Boels:2007qn,Boels:2007gv}.  The twistor actions are compatible with, but lack, the $--+$ amplitude at tree level and the all-plus amplitude at one-loop.  These amplitudes are not constructible from MHV diagrams but are also well localized in twistor space \cite{Witten:2003nn,Cachazo:2004kj}.  

Transformations have been found converting the standard Yang-Mills or QCD Lagrangians into Lagrangians whose tree-level expansion reproduces the CSW rules \cite{Mansfield:2005yd,Ettle:2008ey}, relying on an expansion in lightcone gauge.  One would expect that the transformed Lagrangian then includes the exceptional interaction vertices ($--+$ amplitude at tree level and the all-plus amplitude at one-loop) in some form.  Indeed, they show up as counterterms after suitable regulators are introduced \cite{Ettle:2007qc,Brandhuber:2007vm}.  However, the regulators either break Lorentz invariance or violate the equivalence theorem and exclude supersymmetry.

\section{Recursive constructions of coefficients or rational terms}

BCFW-type on-shell recursion relations can be constructed for parts of loop amplitudes lacking branch cuts.  Most obviously, the rational terms of one-loop amplitudes seem suitable for such treatment.  It is also interesting to consider recursion relations constructed for the {\em coefficients} of the master integrals, as these are also purely rational functions.  

Indeed, if we calculate the coefficients from cut integrals, then they are naturally related to the tree amplitudes on either side of the cut.  Recursion relations for these tree amplitudes lead to recursion relations for the coefficients.  The difference is of course that we now consider shifts involving the loop momentum, and therefore the pole structure is somewhat more subtle.

Recursion relations are derived from factorization properties.
Study of collinear and multiparticle factorization limits shows that it is safe to apply a BCFW shift on two legs attached to the same vertex of the master integral, as long as this shift is valid within that particular tree amplitude \cite{Bern:2005hh}.  With such a shift, kinematic poles involving the loop momentum remain unaffected.  Recursion relations of this type have been written explicitly for gluon amplitudes in a split helicity configuration, with either an $\N=1$ chiral multiplet or a complex scalar circulating in the loop.  

\bigskip

On-shell recursion relations can be used to compute the rational parts of massless one-loop amplitudes \cite{Bern:2005hs,Bern:2005cq,Berger:2006vq,Dunbar:2010xk} (see \cite{Bern:2007dw,Berger:2009zb} for reviews).  We can get the coefficients of master integrals purely from four-dimensional cuts, and find the remaining parts with the residue theorem directly in four-dimensional kinematics as well.  ``Rational term'' could be an ambiguous name, since the master integrals contain some rational terms along with their logarithms; here we define it as the remaining four-dimensional part of the amplitude when all scalar integrals are set to zero, as in (\ref{basis-exp-rational}):
\bea
{\cal R} = 
A^{\rm 1-loop} - \sum_n \sum_{{\bf K}=\{K_1,\ldots,K_n\}} 
\left. c_n({\bf K}) \right|_{\eps=0}
 I_n({\bf K}) 
\Label{ggup}
\eea
We assume that the coefficients of the integrals have been computed first.

Now, examine the poles of this rational part.  They include not only the expected physical poles, but also spurious poles that will be cancelled in the sum with the cut part.  
An example of this kind of unphysical singularity can be seen in the denominator of the two-mass triangle; one can verify that the singular behavior is rational.
These spurious poles can be simple poles or double poles. 
Moreover, it is sometimes difficult to find a momentum shift such that the amplitude vanishes in the limit that the shift parameter $z$ goes to infinity.  Recursion relations for rational parts of one-loop amplitudes have to deal with both of these problems.  

The spurious poles can be canceled explicitly by redefining the rational term.  In equation (\ref{ggup}) above, we have, schematically for an $m$-point amplitude,
\bea
A_m^{\rm 1-loop} =  {\cal C}_m +{\cal R}_m 
\eea
where ${\cal C}_m$ is the cut-containing part with master integrals and ${\cal R}_m$ is the rational term.   There is a systematic (but not unique) way to ``complete'' the cut part by adding some rational terms so that the spurious poles are cancelled, starting with two-mass triangle integrals and including higher-dimensional box and triangle integrals as necessary \cite{Berger:2006ci,Campbell:1996zw}.  If these extra terms are represented by $\widehat{\cal CR}_m$, then the cut completion is 
\bea
\widehat{\cal C}_m = {\cal C}_m + \widehat{\cal CR}_m
\eea
and we recast the rational part as
\bea
\widehat{\cal R}_m = {\cal R}_m - \widehat{\cal CR}_m.
\eea
Now a recursion relation can be constructed for $\widehat{\cal R}_m$ from knowledge of its physical poles.  In a given factorization limit, there are three types of contributions, corresponding to the location of the original loop integral.  
The recursion relation takes the form
\bea
\hspace{-1in}
\widehat{\cal R}_m = \sum_{{\rm poles}~\alpha}
\left(
\widehat{\cal R}_L(z_\alpha){i \over P_\alpha^2} A_R(z_\alpha)
+ A_L(z_\alpha){i \over P_\alpha^2}\widehat{\cal R}_R(z_\alpha)
+  A_L(z_\alpha){i {\cal F}_\alpha \over P_\alpha^2} A_R(z_\alpha)
\right).
\Label{1-loop-3-term}
\eea
Here, $A_L$ and $A_R$ are lower-point tree amplitudes; $\widehat{\cal R}_L$ and $\widehat{\cal R}_R$ are rational parts of lower-point one-loop amplitudes; and the ``factorization function'' ${\cal F}$ is derived from diagrams where the loop is a correction to the propagator that is going on shell \cite{Berger:2006ci}.

This three-term recursion relation is still not completely correct, because there are generally also boundary terms coming from the non-vanishing of the amplitude as $z \to \infty$.  And when it is possible to select a shift such that the amplitude does vanish at infinity, then there is a similar issue in the recursion relation itself: along with the three recursive terms in equation (\ref{1-loop-3-term}), there are extra terms in the residues themselves that do not fit the recursive pattern. 

The resolution is to combine information from two different shifts.  First, use a primary shift, under which the residues are purely recursive as in equation (\ref{1-loop-3-term}).  Then look at the boundary term, described by the large-$z$ behavior of the amplitude and denoted by ${\rm Inf}~A_m^{\rm 1-loop} $.  This boundary term can be constructed from a recursion relation using an auxiliary shift.  The auxiliary shift may have non-recursive terms in its residues, but they can be dealt with as long as they vanish in the large $z$ limit.  We also need to compute ${\rm Inf}~\widehat{\cal C}_m$, the large $z$ behavior of the cut-completed terms, under {\em both} shifts separately, and take care to avoid double-counting the residues at physical poles in the cut-completed and rational parts.

\section{Survey of recently published one-loop amplitudes}

 Since the introduction of the unitarity method in 1994, many new analytic, explicit expressions for one-loop amplitudes have become available.  Here we list those that are the most complete (all helicity configurations and rational terms). 

Gluon amplitudes in $\N=4$ SYM are known explicitly for MHV \cite{Bern:1994zx} and NMHV \cite{Bern:2004bt} configurations, which is enough to give all results through seven external legs \cite{Bern:2004ky}.  

In QCD, partial results for five-parton amplitudes \cite{Bern:1993mq,Kunszt:1994tq} were completed with the unitarity method and summarized in \cite{Bern:1994fz}

The amplitudes for $e^+ e^- \to$ four partons (also necessary for W/Z/Drell-Yan production in association with two jets at hadron colliders, or three-jet production in deep inelastic scattering) have been given in \cite{Bern:1997sc}, which also featured the used of a triple cut to find a three-mass triangle coefficient.

The components of the six-gluon amplitude \cite{Mahlon:1993si,Bern:1993qk,Bern:1994zx,Bern:1994cg,Bidder:2004tx,Bedford:2004nh,Britto:2005ha,Bern:2005cq,Bern:2005hh,Britto:2006sj,Berger:2006ci,Berger:2006vq,Xiao:2006vt} have been summarized in \cite{Dunbar:2008zz,Dunbar:2009uk} with an accompanying {\tt Mathematica} code.  The results have been cross-checked with a semi-numerical derivation \cite{Ellis:2006ss}.

The six-photon amplitude in QED has been found from both multi-cut and form factor decomposition methods and cross-checked in \cite{Binoth:2007ca}.  The result was  given for scalar QED and $\N=1$ QED by cut methods in \cite{Bernicot:2008nd}.  Both analytic forms confirmed the numerical results of \cite{Nagy:2006xy}.

The components of the Higgs + four parton amplitudes have been found in the limit of large top quark mass \cite{Badger:2006us,Berger:2006sh,Badger:2007si,Glover:2008ffa,Dixon:2009uk,Badger:2009hw,Badger:2009vh}.  The analytic results give a fast (10 ms) evaluation of the full cross section, and the  semi-numerical result \cite{Ellis:2005qe,Campbell:2006xx} has been confirmed.
 
Amplitudes featuring massive fermions are a bit more rare.  Four-gluon scattering via a massive quark loop was fully calculated using cut methods in \cite{Bern:1995db}.

Formulas for the $ t\bar{t}gg$  amplitude have been found from cut methods and described  in \cite{Badger:2008za,Badger:2010wf} and have been checked against the  earlier (less compact) results of \cite{Korner:2002hy,Anastasiou:2008vd}.

Lately, the calculation of the  $W$-boson mediated process $0 \to d\bar{u}Q\bar{Q}\bar{\ell}\ell$ \cite{Badger:2010mg} has been presented, an  analytic version of \cite{FebresCordero:2006sj} and including the spinor correlations present in W-boson decay.

\section{Beyond one loop}

Unitarity and recursive methods have been applied to compute multiloop amplitudes, but almost only in theories of maximal supersymmetry, which merit a separate discussion.

Maximal supersymmetry implies many internal cancellations and hence a relatively simple integral structure.  Moreover, $\N=4$ SYM and $\N=8$ supergravity possess symmetries even beyond those directly implied by gauge symmetry and supersymmetry, simplifying the results even further.

It is worth noting that even in maximally supersymmetric theories, multiloop calculations have been completed only for MHV configurations.  No one should be astonished, then, that progress with reduced supersymmetry is taking more time.  

With on-shell methods, there are several obstacles to address to proceed from one loops to two or more.  First, some convenient simplifications within QCD do not carry over.  At one-loop, we could apply a color decomposition and calculate color-stripped ``partial amplitudes.''  The subleading-color contributions were related by permutation identities to the leading-color part \cite{Bern:1990ux,Bern:1994zx}.  At two loops, we can start with a color decomposition, but the subleading contributions need to be computed separately.  

In QCD, 
one-loop gluon amplitude calculations have routinely proceeded according to a supersymmetry decomposition: first, complete the internal gluon with a full $\N=4$ supermultiplet, and compute the amplitude in the $\N=4$ theory; then, remove the new fermion contributions again by subtracting four times the $\N=1$ amplitude with a chiral multiplet in the loop; finally, cancel all the scalar by explicitly adding a nonsupersymmetric amplitude with a complex scalar in the loop.  The decomposition is written schematically as
\bea
A_{\rm 1-loop}^{\N=0~{\rm gluon}} = A_{\rm 1-loop}^{\N=4} - 4 A_{\rm 1-loop}^{\N=1{\rm ~chiral}} + A_{\rm 1-loop}^{\rm scalar}.
\eea
The underlying principle is that supersymmetric amplitudes are simpler due to internal cancellations (for example, diagrams loop amplitudes in $\N=4$ SYM have no internal triangle subdiagrams), and that a nonsupersymmetric amplitude with a complex scalar loop is simpler than with a gluon loop, because there are no spin degrees of freedom.  At two loops, there is no relation of this kind.  One can hope to get some insight by starting with a calculation with internal scalars, which can help in formulating an ansatz for the full solution \cite{Bern:2000dn}, but it is not a systematic approach.  

Even more seriously, in general theories, the master integrals are more numerous and less explicit; and a unitarity cut of an amplitude is unavoidably $D$-dimensional, typically with internal spins.  The four-dimensional spinor-helicity formalism has been tremendously useful in writing amplitudes analytically.  Higher-dimensional spinor formalisms have been explored \cite{Cheung:2009dc,Dennen:2009vk,Boels:2009bv,Bern:2010qa,Brandhuber:2010mm} but still need further attention.  Continuations to larger {\it integer} values of $D$ are enough to fix all rational terms of one-loop amplitudes \cite{Giele:2008ve} and certainly help in defining the tree-level input for cut-based methods.

Two-loop master integrals can be generated for any specific process by the Laporta algorithm \cite{Laporta:2000dc,Laporta:2001dd}, which is based on integration-by-parts identities relating various integrals in systems of (large numbers of) linear equations.  Programs implementing the Laporta algorithm efficiently are available \cite{Anastasiou:2004vj,Smirnov:2008iw,Studerus:2009ye}.  Recently a framework has been outline for reducing two-loop integrals to a finite basis avoiding doubled propagators, whose singularities are more difficult to control \cite{Gluza:2010ws}.  It still remains to list them explicitly, check independence, and evaluate them.

The operations that have been proposed to construct one-loop amplitudes from their single cuts using  different propagator prescriptions  \cite{CaronHuot:2010zt,Catani:2008xa} can be extended to higher-loop integrals \cite{Bierenbaum:2010cy}, since there is no need for a basis of master integrals, and the information regarding integration can be said to be carried in the propagator prescriptions.

In QCD, unitarity methods have been applied to compute the 4-gluon amplitude to two-loop order \cite{Bern:2000dn}, as well as the two-loop splitting amplitudes (universal functions governing collinear behavior) for  $g \to gg$ \cite{Bern:2004cz} and  $q \to qg$ and $g \to q\bar q$ \cite{Badger:2004uk}.
Seven master integrals were needed, of which four vanished through $\ord(\eps)$ and two others were products of one-loop integrals.  In principle there could have been two-loop planar integrals that had no cuts in two-particle channels; fortunately, these did did not appear in this particular calculation, and so it sufficed to look at cuts in the physical channels.
The corresponding amplitude with internal scalars were found first as an exercise.  

\section*{Acknowledgments}

I wish to thank C. Anastasiou, E. Buchbinder, F. Cachazo, B. Feng, Z. Kunszt, P. Mastrolia, E. Mirabella and G. Yang for collaboration on topics presented in this review, and S. Badger, R. Boels, D. A. Kosower, D. Skinner and E. Witten for additional illuminating discussions.   My work is supported by the {\it Agence Nationale de la Recherche} under grant ANR-09-CEXC-009-01.

\section*{References}

\bibliographystyle{unsrt}
\bibliography{references}

\end{document}